\documentclass[final,5p,times,twocolumn]{elsarticle}

\usepackage{graphicx}

\usepackage{amssymb}

 \usepackage{lineno}

\newcommand{\neqcm}{\ensuremath{\mathrm{n}_{\mathrm{eq}}/\mathrm{cm}^2}}

\journal{Nuclear Instruments and Methods A}
\begin{document}

\begin{frontmatter}



\title{Development of n-in-p pixel modules for the ATLAS Upgrade at HL-LHC}



\author{A. Macchiolo}
\author{R. Nisius}
\author{N. Savic}
\author{S. Terzo}

\address{Max-Planck-Institut for Physics, F\"{o}hringer Ring 6, D-80805 Munich, Germany}

\begin{abstract}

Thin planar pixel modules are promising candidates to instrument the inner layers of the new ATLAS pixel detector for HL-LHC,
 thanks to the reduced contribution to the material budget and their high charge collection efficiency after irradiation. 
100-200 $\mu$m thick sensors, interconnected to FE-I4 read-out chips, have been characterized with radioactive sources and beam tests
 at the CERN-SPS and DESY. The results of these measurements are reported for devices before and after irradiation up to a fluence of  $14\times10^{15}$ \neqcm{}. The charge collection and tracking efficiency of the different sensor thicknesses are compared. The outlook for future planar pixel sensor production is discussed, with a focus on sensor design with the pixel pitches (50x50 and 25x100 $\mu$m$^2$) foreseen for the RD53 Collaboration read-out chip in 65 nm CMOS technology. An optimization of the biasing structures in the pixel cells is required to avoid the hit efficiency loss presently observed in the punch-through region after irradiation. For this purpose the performance of different layouts have been compared in FE-I4 compatible sensors at various fluence levels by using beam test data. Highly segmented sensors will represent a challenge for the tracking in the forward region of the pixel system at HL-LHC. In order to reproduce the performance of 50x50 $\mu$m$^2$ pixels at high pseudo-rapidity values, FE-I4 compatible planar pixel sensors have been studied before and after irradiation in beam tests at high incidence angle (80$^\circ$) with respect to the short pixel direction. Results on cluster shapes, charge collection and hit efficiency will be shown.

\end{abstract}


\end{frontmatter}


\section{Introduction}
In this paper, different designs of n-in-p planar hybrid pixel modules are investigated and compared.
The R\&D activity is carried out in view of the upgrade of the ATLAS pixel system
for the  High Luminosity phase of the LHC (HL-LHC) \cite{HL-LHC}, foreseen to start around 2025.
The number of pile-up events per bunch crossing expected at the HL-LHC is 140-200 \cite{HILUMI1, HILUMI2}. To keep 
the pixel occupancy at an acceptable level, smaller pixel cell dimensions are foreseen with respect to the 
ones presently implemented in the FE-I3 chip  (50 $\mu$m x 400 $\mu$m) and in the
FE-I4 chip (50 $\mu$m x 250 $\mu$m), developed for the ATLAS Insertable B-Layer (IBL) \cite{IBL}.
The new readout chips for the ATLAS and CMS pixel systems at HL-LHC are being developed by the CERN RD53 Collaboration
\cite{RD53} with a pitch of 50 $\mu$m x 50 $\mu$m in the 65 nm CMOS technology.
The feasibility  of employing thin planar pixel detectors for the inner layers 
of the upgraded ATLAS pixel system is being investigated, comparing the performance
of sensors in the thickness range of 100-200 $\mu$m.  Using the SOI technology, n-in-p sensors with a full thickness of 100 and 200 $\mu$m  were produced on 6"  wafers at VTT in the framework of a Multi-project Wafer run (MPW) \cite{Kalliopuska}.

This production also implements activated edges to reduce the inactive area of these devices \cite{VTT1} \cite {VTT2} and 
the performance in terms of charge collection at the edges has been presented elsewhere \cite{IWORID}.
200  $\mu$m thick sensors were also produced at CIS, on 4" wafers, with a standard guard ring structure and
an inactive width of 450 $\mu$m \cite{stefano_thesis}.

The sensors have been interconnected 
by bump-bonding to FE-I4 chips and characterized by means of radioactive sources in the laboratory
and beam test experiment with 4 GeV electrons at DESY and 120 GeV pions at CERN-SPS.

\section{Characterization of sensors with different thickness after irradiation}
\begin{figure}[ht]
\centering
\includegraphics[width=\columnwidth]{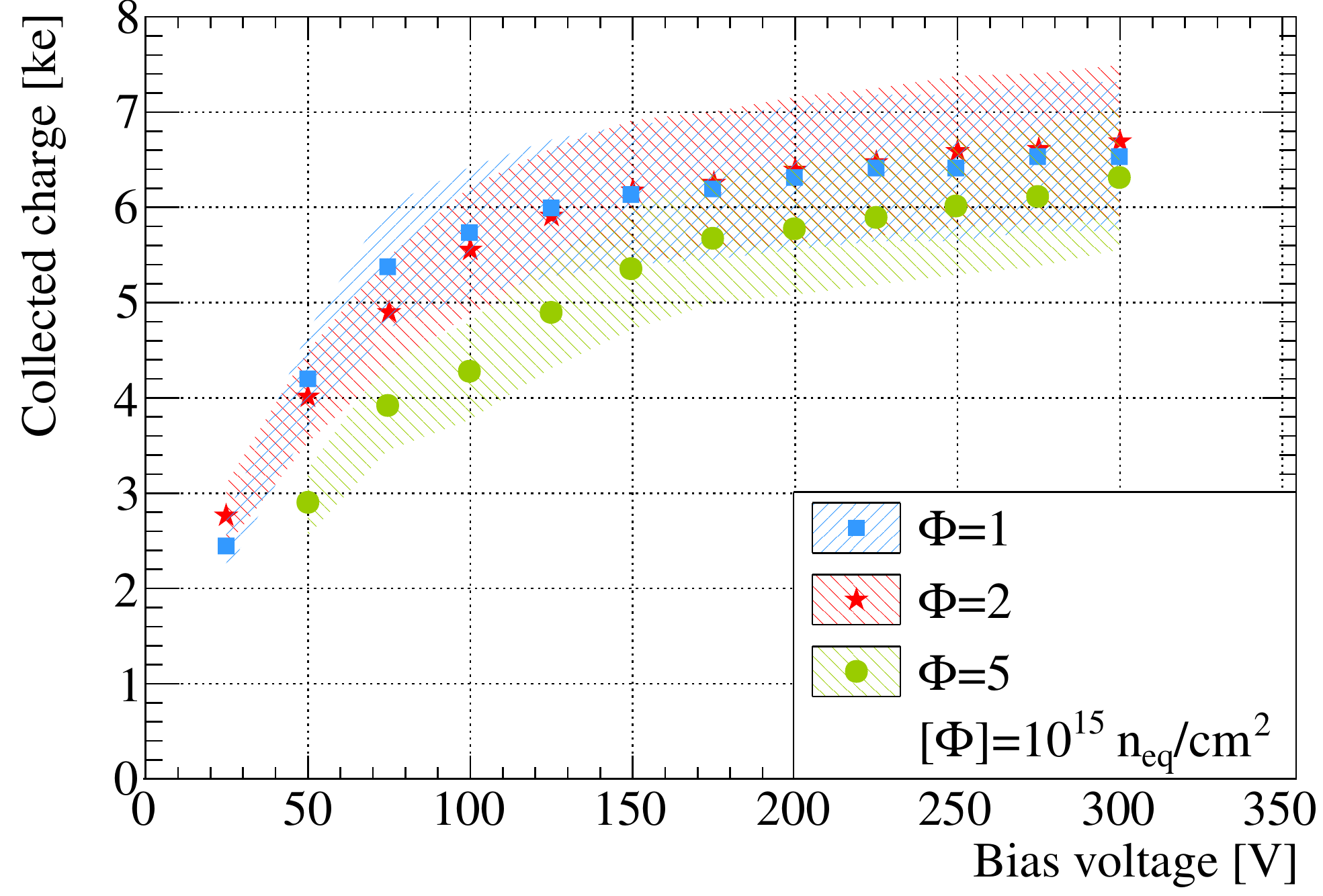}
\caption{Charge collection of 100 $\mu$m thick sensors as a function of the bias voltage for different fluences}
\label{fig:CCE_100}
\end{figure}

\begin{figure}[ht]
\centering
\includegraphics[width=\columnwidth]{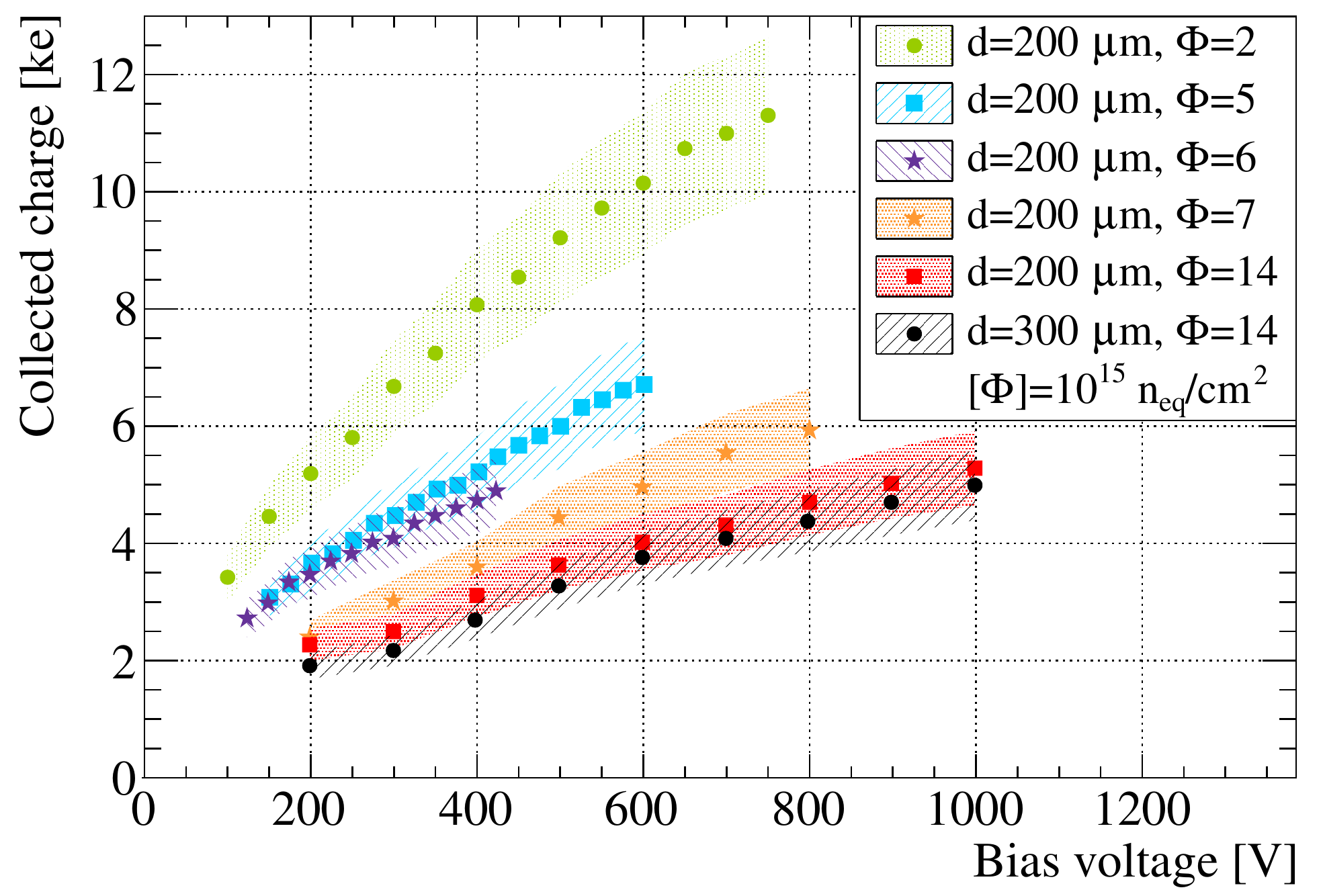}
\caption{Charge collection of 200 and 300 $\mu$m  thick sensors as a function of the bias voltage for different fluences}
\label{fig:CCE_200}
\end{figure}

The charge collection properties were  measured by means of Strontium source scans 
using the ATLAS USBpix read-out system, developed by the University of Bonn \cite{USBPix}. The pixel modules
were investigated before and after irradiation,
up to a fluence  of $5\times10^{15}$ \neqcm  for the  100 $\mu$m thick sensors, corresponding to the integrated 
fluence at the end of the life-time for the second pixel layer at HL-LHC and for the 
 200 $\mu$m thick sensors up to a fluence of $14\times10^{15}$ \neqcm, 
the expected value for the inner layer after ten years of operation at HL-LHC.
Fig.\ref{fig:CCE_100} shows the collected charge for the 100 $\mu$m \, thick sensors, as a function of the applied bias voltage for different
fluences. It can be observed that the collected charge starts to saturate at moderate  bias voltages, between 200 and 300V, 
in the full fluence range explored, to values that are very close to those expected for not irradiated detectors, around 6500-7000
electrons \cite{thesis_philipp}.
Fig.\ref{fig:CCE_200} shows the results of the charge collection for the 200 $\mu$m thick sensors and of 300 $\mu$m thick sensors
at the highest fluence only. After $2\times10^{15}$ \neqcm,
no  saturation of the charge is observed. At the maximum fluence
of $14\times10^{15}$ \neqcm, the collected charge of the 200 and 300 $\mu$m thick sensors becomes very similar 
and at the highest measured voltage of 1000V is between 5 ke and 5.5 ke.

\begin{figure}[hbt] 
\centering 
\includegraphics[width=\columnwidth,keepaspectratio]{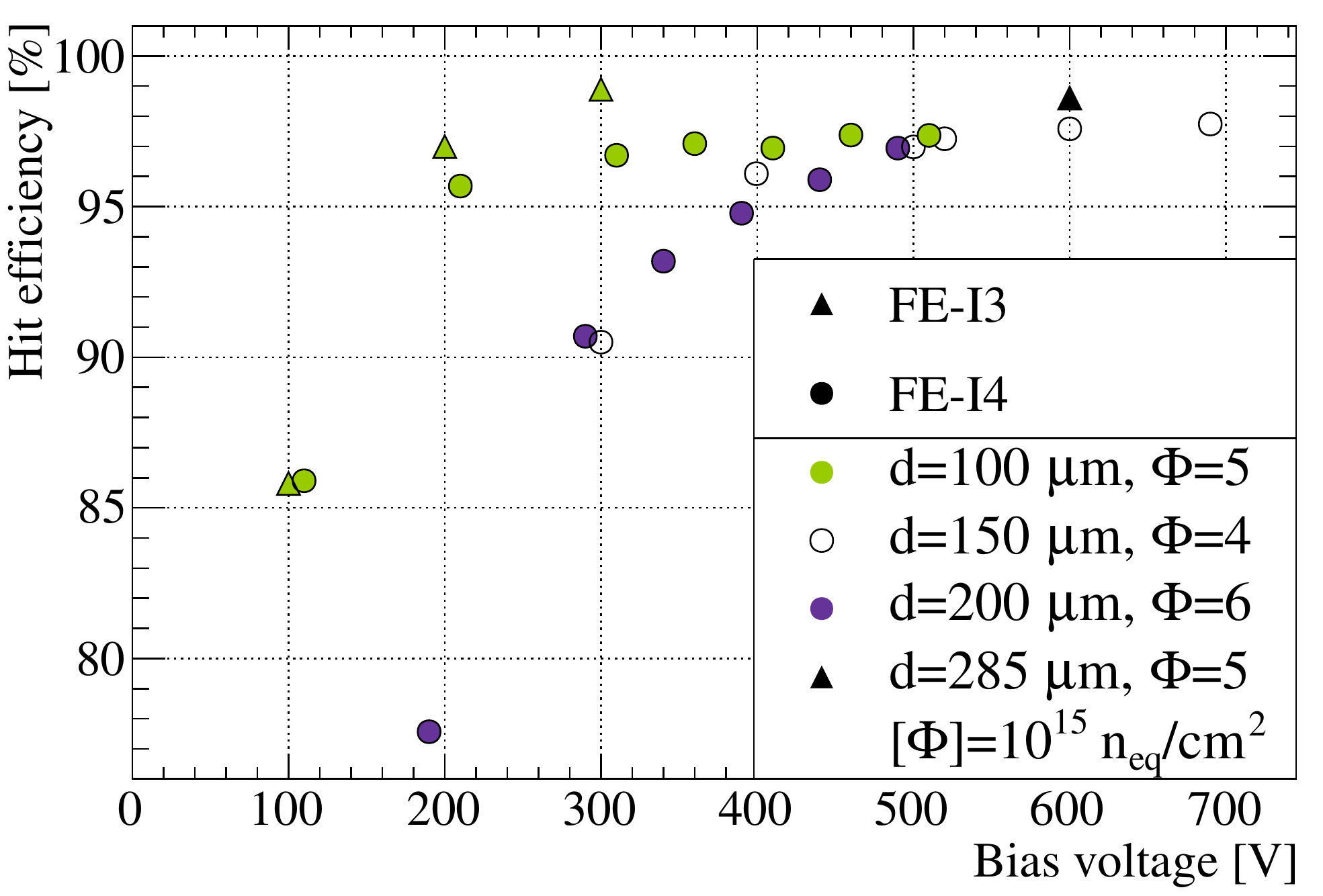}
\caption{Hit efficiency as a function of the bias voltage  for pixel modules of different thicknesses irradiated
to a fluence  between 4 and $6\times10^{15}$ \neqcm. The modules were operated at a threshold of 1600 e.}
\label{fig:hit_eff}
\end{figure}

\begin{figure}[hbt] 
\centering 
\includegraphics[width=\columnwidth,keepaspectratio]{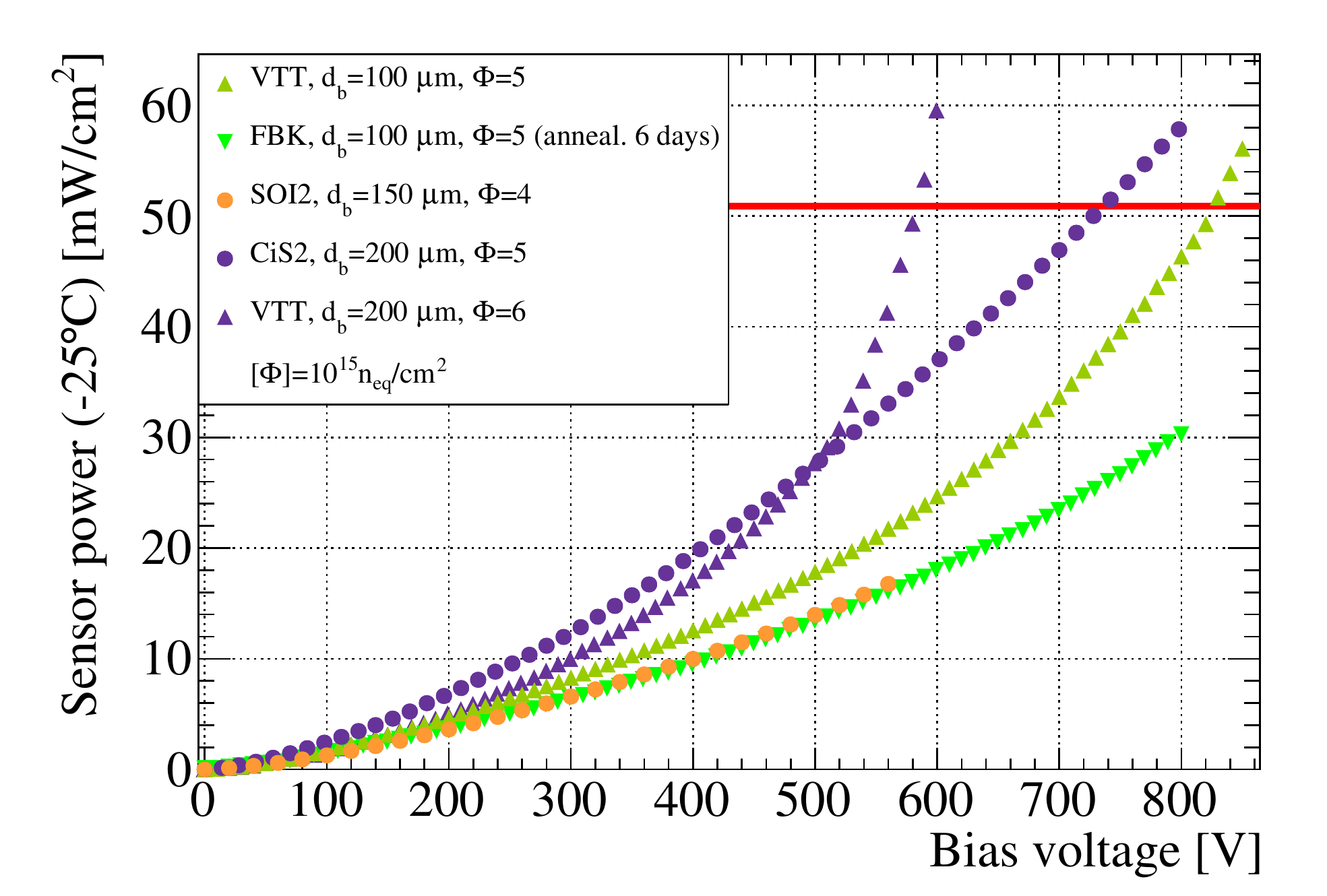}
\caption{Power density as a function of the bias voltage for sensors irradiated between 4 and $6\times10^{15}$ \neqcm
estimated at a temperature of T=-25 $^\circ$C. The horizontal red line indicates
the thermal runaway point as calculated in \cite{stefano_thesis}.}
\label{fig:powerdissipation}
\end{figure}

FE-I3 and FE-I4  modules of different thickness were also investigated in beam tests by using telescopes of the EUDET 
family \cite{EUDET}. The irradiated modules were cooled with dry ice inside a box designed on purpose. With this setup typical temperatures between -50$^\circ$C and -40$^\circ$C are obtained during operations. Recent studies have shown the independence of the 
collected charge in a temperature range between -50$^\circ$C and -25$^\circ$C \cite{RD50_santander} and between 
-50$^\circ$C and -40$^\circ$C for the hit efficiency \cite{RD50_dec2015}.
Figure \ref{fig:hit_eff} summarizes the results 
of the hit efficiency in a fluence range between 4 and $6\times10^{15}$ \neqcm. The 150 $\mu$m thick sensors
were obtained from an older production at MPG-HLL on SOI 6" wafers while the 285 $\mu$m thick sensors 
were produced on 4" wafers at CIS, as described in more details in \cite{thesis_philipp}.
For all the values of the hit efficiencies quoted in the following, the dominant source of uncertainty is systematic and evaluated
to be 0.3$\%$, as explained in \cite{test-beam}.
After an irradiation fluence between 4 and $6\times10^{15}$ \neqcm, FE-I4 modules with 150 and 200 $\mu$m thick sensors show similar performances
reaching a hit efficiency of about 97$\%$ at V$_{bias}$=500V, while the same module type with a 100$\mu$m
thick sensor starts to saturate to this value of the hit efficiency already at a bias voltage of 300V. These results
suggest that a lower operational bias voltage is possible for thinner sensors, and a reduced power dissipation 
at these fluence levels. 
Fig.\ref{fig:powerdissipation} shows the expected sensor power density for these devices at a temperature of -25 $^\circ$C,
 at which it is foreseen that the sensors will be operated at HL-LHC, as a function of the applied bias voltage for different values of the sensor thickness. 
A high power consumption requires sufficient performance of the cooling system to dissipate the heat and avoid the thermal runaway of the sensor.
Since this effect limits the achievable operational voltage of the modules,
 it has to be considered together with the hit efficiency results to determine the practical performance of the different sensor thicknesses.
The power per area calculated for the thinner sensors of 100 and 150 $\mu$m at the
optimal operational voltage of 300 and 500V estimated from the hit efficiency measurements in
Fig.\ref{fig:hit_eff}, is respectively 8 and 3.5  times lower than the runaway point of the worst case considered.
The runaway point was calculated with the parameters descibed in \cite{stefano_thesis}, except for the higher chip power 
dissipation of 0.74 W/cm$^2$ assumed for the RD53 Collaboration read-out chip in 65 nm CMOS technology.

\section{Optimization of the pixel cell design}
\begin{figure}[hbt] 
\centering 
\includegraphics[width=\columnwidth,keepaspectratio]{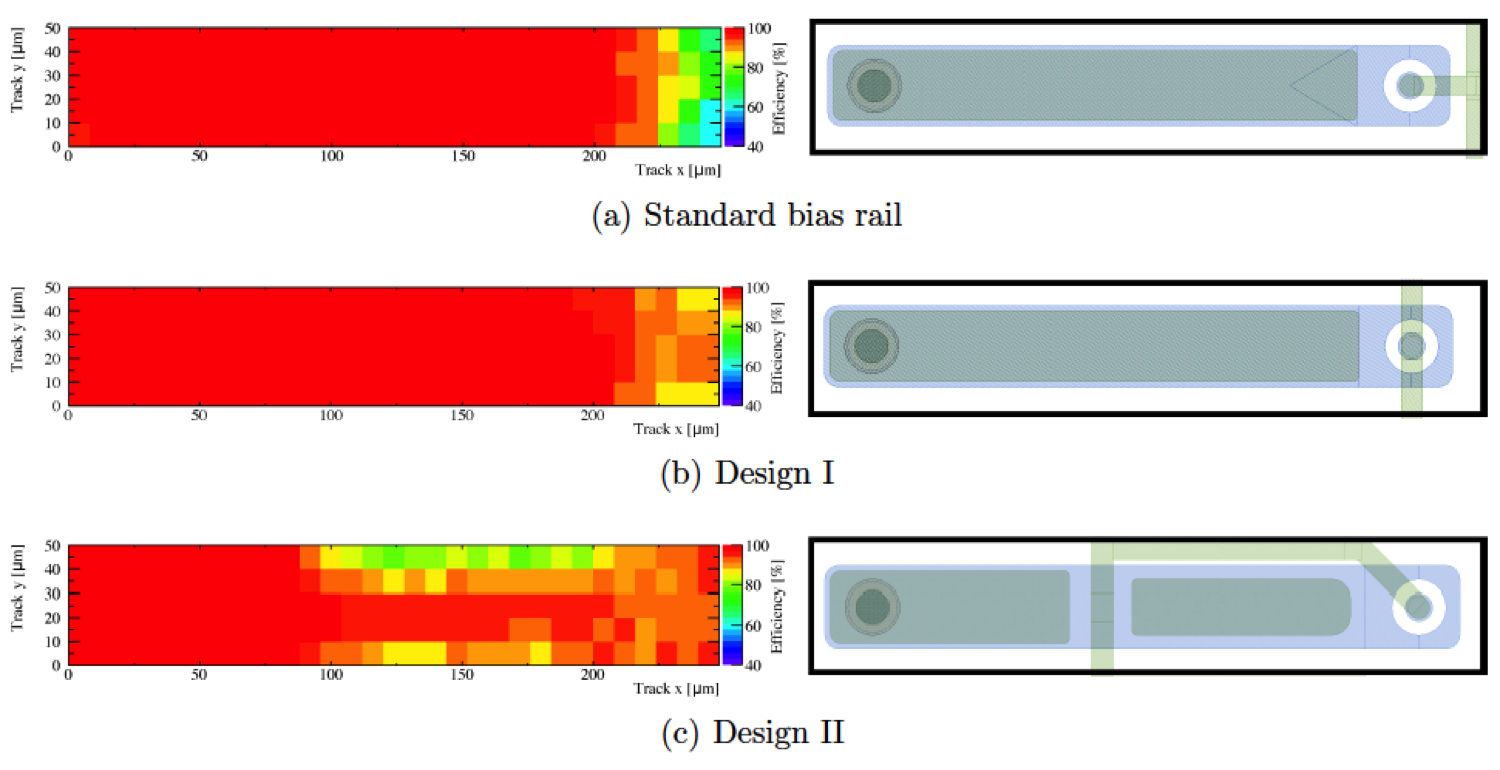}
\caption{Hit efficiency projected onto a pixel cell for different single bias dot designs after an irradiation
 fluence of $3\times10^{15}$ \neqcm. The module was operated at 800V.}
\label{fig:bias_dot}
\end{figure}

Figure \ref{fig:hit_eff}  shows that at equal fluence and thickness, FE-I3 modules yield a hit efficiency 1$\%$  higher with respect
to the FE-I4.
This is due to the lower fraction of area that the biasing structures occupy in the pixel cell, since the punch-through
dot and the bias rail are implemented with the same design and the FE-I3 pixel lenght is 400 $\mu$m compared with the 250 $\mu$m
in the FE-I4 case. It has found that these elements induce a decrease of the hit efficiency in the pixel cell \cite{stefano_thesis}.
Given the reduced pitch for the future pixel read-out chips, an optimization of the biasing structures is 
mandatory, to avoid a large loss of hit effiiciency after irradiation, especially in the central pseudo-rapidity region.\\
Different designs of the bias dot and rail were thus implemented in two different pixel sensors, both compatible with the FE-I4 chip,
 of a recent n-in-p production carried out at CIS on 6" wafers, with a thickness of 270 $\mu$m.
 The first one includes three designs, repeated in every successive group of 30 rows, 
 with a single bias dot for every pixel: the standard design (Fig. \ref{fig:bias_dot} (a)) with the rail at the center between the short side of two neighbouring pixels, a slightly modified version with the rail running over the bias dot (b) and 
the last one (c), with the rail over the central part of the cell. After irradiation at a fluence of $3\times10^{15}$ \neqcm, the hit efficiencies of the different
designs have been compared. The highest value of 98.7$\%$ has been found for the geometry (b), while the standard
geometry yields an efficiency of 97.7$\%$. The layout in Fig. \ref{fig:bias_dot} (c) shows the lowest efficiency, especially in correspondence
of the horizontal lines of the bias rail that runs in the inter-pixel region. It can instead be observed that where the bias rail is superimposed to the 
pixel implant, both in (b) and in (c), the aluminum line does not induce any efficiency loss because the effects on the electric field shapes are then screened by the pixel implant.

\begin{figure}[hbt] 
\centering 
\includegraphics[width=\columnwidth,keepaspectratio]{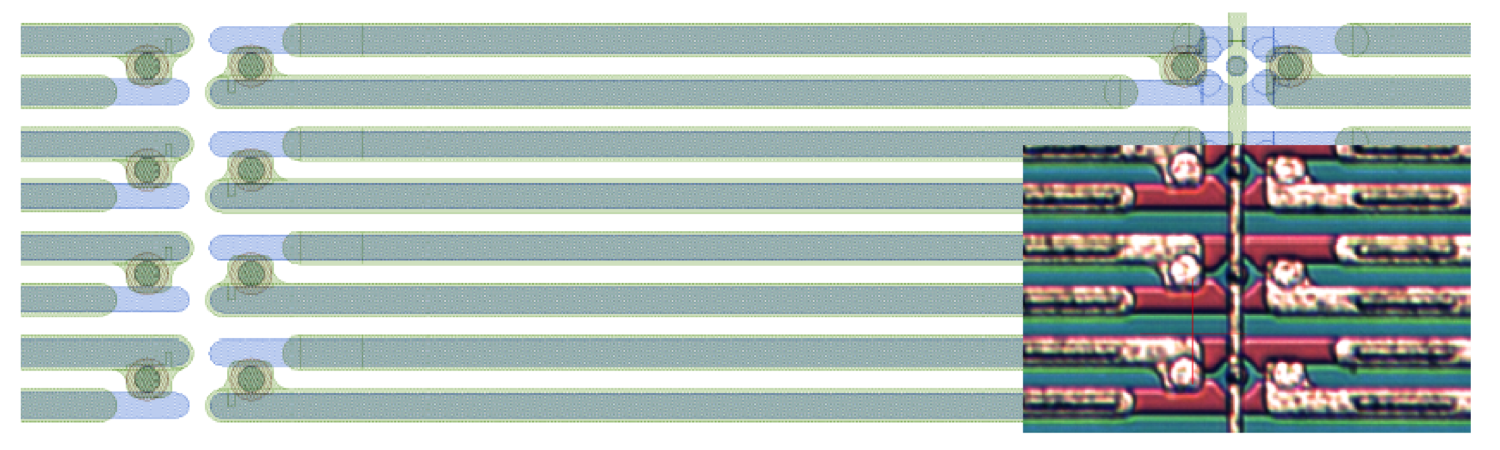}
\caption{ Layout of the 25 x500 $\mu$m$^2$ FE-I4 compatible sensor design with an external punch-through common to four pixel cells. The insert is a photograph of the produced device superimposed to the schematic design of the sensor.}
\label{fig:external_dot}
\end{figure}

\begin{figure}[hbt] 
\centering 
\includegraphics[width=\columnwidth,keepaspectratio]{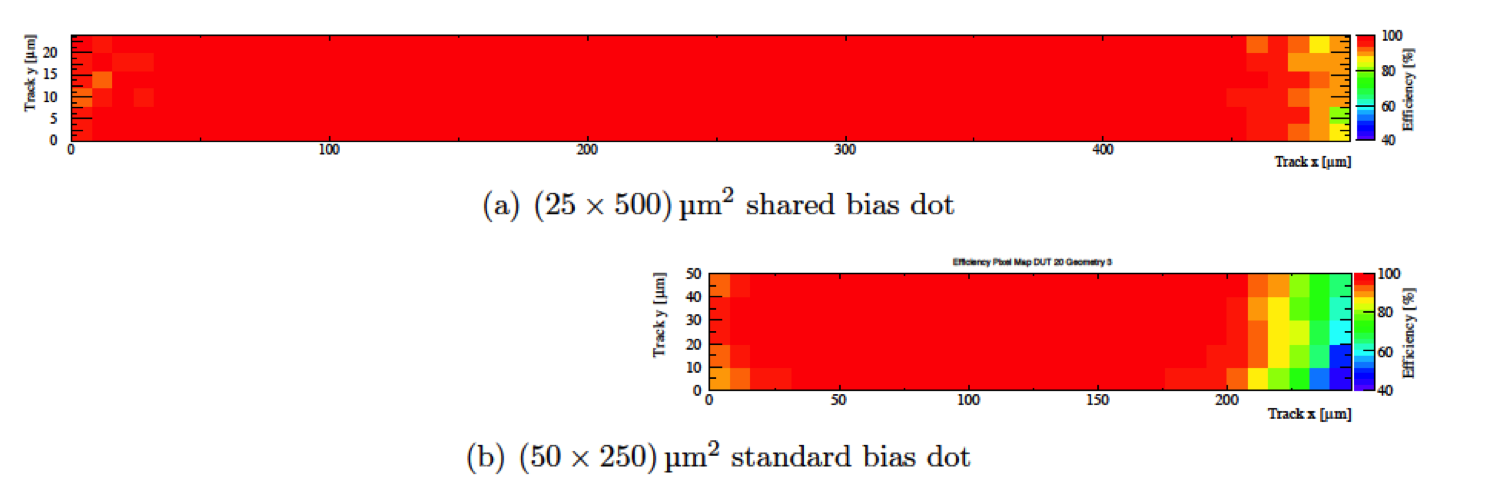}
\caption{Comparison of the hit efficiency over the pixel cell for the FE-I4 modules of the CIS3
production irradiated to a fluence of $3\times10^{15}$ \neqcm with different bias structures. (a) shows
the hit efficiency for the module with 25 x500 $\mu$m$^2$ pitch and one common bias dot shared 
among four pixels. (b) shows the hit efficiency for the standard FE-I4 design with an internal punch-through dot
and a pitch of 50 x 250 $\mu$m$^2$.}
\label{fig:hiteff_commondot}
\end{figure}

 Higher hit efficiencies are obtained by implementing an external punch-through dot, common to four pixels, as shown by Fig. \ref{fig:external_dot} and \ref{fig:hiteff_commondot}.
In this case the sensor pitch is 25 $\mu$m x 500 $\mu$m, still compatible with a FE-I4 chip.
 The overall area where a lower hit efficiency  is observed is clearly reduced with respect
to the standard FE-I4 design. The hit efficiency reaches 99.4$\%$ at the highest measured voltage
of 500V which is around 2$\%$ higher than the 97.7$\%$ measured for the standard FE-I4 design at the same fluence.

Two new pixel sensor productions at ADVACAM and CIS, with 100 and 150 $\mu$m thickness, have recently been completed and they
include FE-I4 compatible devices with 50 $\mu$m x 250 $\mu$m pitch and the new external punch-through biasing structures.
 It is planned to repeat the hit efficiency measurements
with these sensors in a fluence range up to $10^{16}$ \neqcm \, to confirm the better performance observed with the 
new biasing design with thinner devices.

\section{Performance of n-in-p planar pixels at high $\phi$}

\begin{figure}[hbt] 
\centering 
\includegraphics[width=\columnwidth,keepaspectratio]{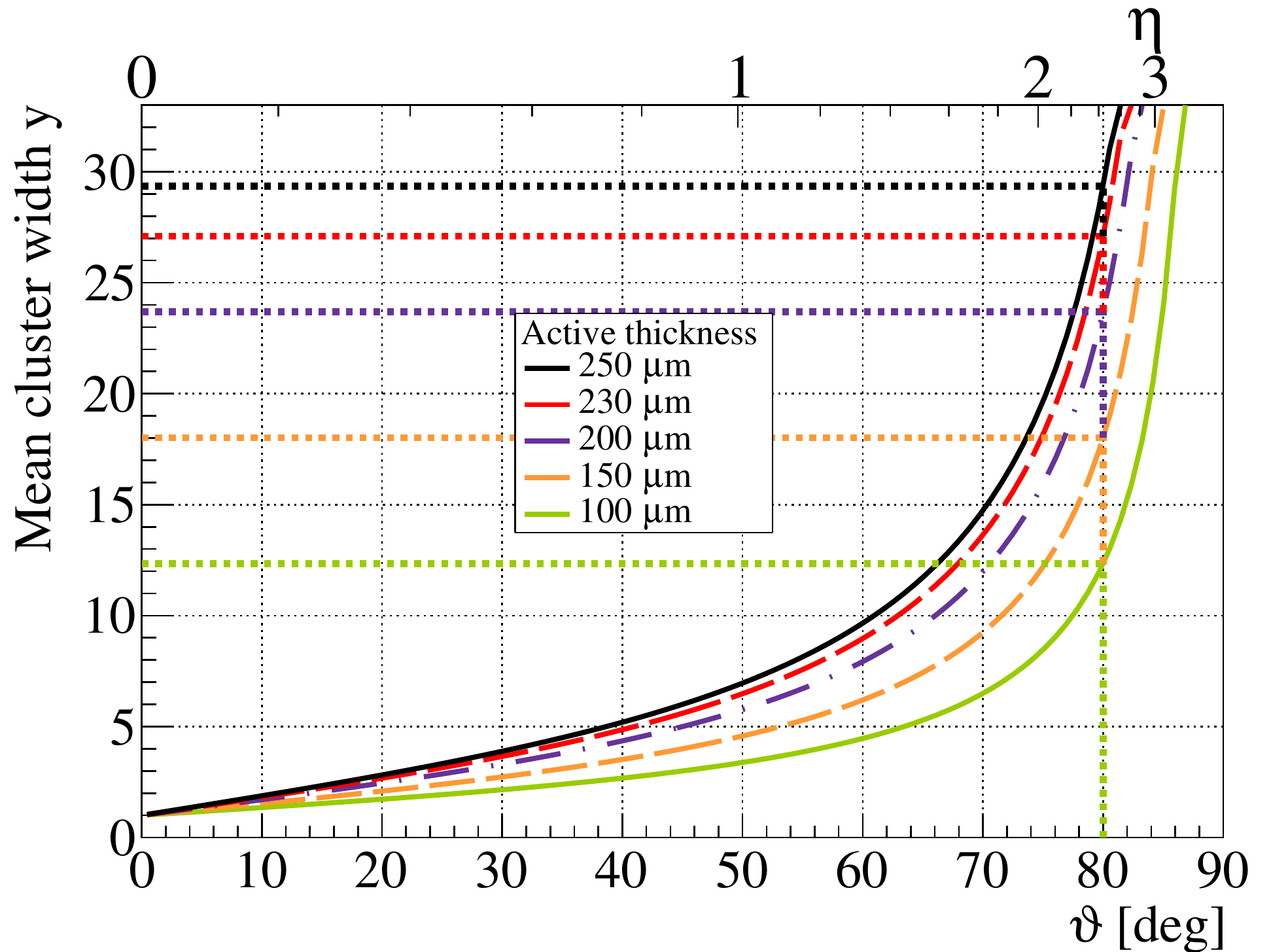}
\caption{Mean cluster width along the short pixel cell side for a FE-I4 module placed at high $\phi$ in the beam, as a function of
beam incidence angle. The relationship is also valid for a pixel sensor with 50 $\mu$m x 50$\mu$m pitch at high $\eta$ with respect
to the beam.}
\label{fig:clustersize}
\end{figure}

\begin{figure}[hbt] 
\centering 
\includegraphics[width=\columnwidth,keepaspectratio]{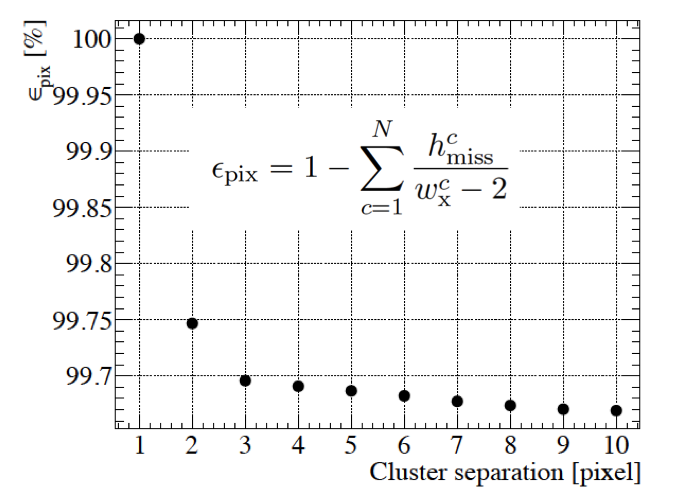}
\caption{Hit efficiency of single pixels as a function of the cluster separation for an FE-I4
module employing a 100 $\mu$m thick sensor and tilted by 80$^\circ$ around its x axis with
respect to the perpendicular beam incidence. The single hit inefficiency is defined as the number of holes ($h^{c}$$_{miss}$) divided by the cluster lenght $w^c_x$,  not including the entrance and exit pixels, since these are 100$\%$ efficient by definition.
The sum is over all the N reconstructed clusters for a given choice of the cluster separation value.
}
\label{fig:singlepixeleff}
\end{figure}

The smaller pixel cell dimensions foreseen at HL-LHC pose a severe challenge also for the tracking
in high pseudo-rapidity regions (high $\eta$) of the new trackers.
To investigate  the hit efficiency for a cell size of 50 $\mu$m x 50$\mu$m at $\eta$=2.5, FE-I4 modules
were investigated with an electron beam at DESY crossing the sensor with an angle of $\phi$= 80$^{0}$ with respect to the pixel surface.
In this set-up  the particles cross the pixel cells along the
50 $\mu$m side, allowing to study the performance of a  50 $\mu$m x 50$\mu$m sensor at high $\eta$.
 Figure \ref{fig:clustersize} shows how the cluster size along $\eta$ is strongly dependent on the sensor
thickness, with thinner sensors yielding the smaller clusters and resulting in the lower pixel occupancy.
The first module employed in this test is a not irradiated device, 100 $\mu$m thick, from the VTT production.
Track reconstruction was not possible for the data sample recorded, and the analysis was performed 
using the hit information of the long clusters corresponding to a single particle traversing the sensor.
The cluster size and hit efficiency were calculated under varying assumptions on the allowed  number of holes
in the cluster (defined as cluster distance). Given the fact that the particle path in silicon is only around 50 $\mu$m long,
with an expected most probable value (MPV) of the collected charge of 3100 e, 
 a dedicated tuning was performed, with a low target threshold of 1000 e. 
Fig.\ref{fig:singlepixeleff} shows the single hit inefficiency as a function of the maximum value of the distance 
 between two pixels allowed while building a cluster (cluster separation).
The single hit inefficiency is defined as the number of holes ($h^{c}$$_{miss}$) divided by the cluster length $w^c_x$, 
not including the entrance and exit pixels, since these are 100$\%$ efficient by definition.
It can be observed that values of the hit efficiency close to 100$\%$ are obtained with this thin sensor module, 
for all the cluster distance hypothesis,
suggesting the feasibility of employing this kind of detectors at high $\eta$ values at HL-LHC.

\begin{figure}[hbt] 
\centering 
\includegraphics[width=\columnwidth,keepaspectratio]{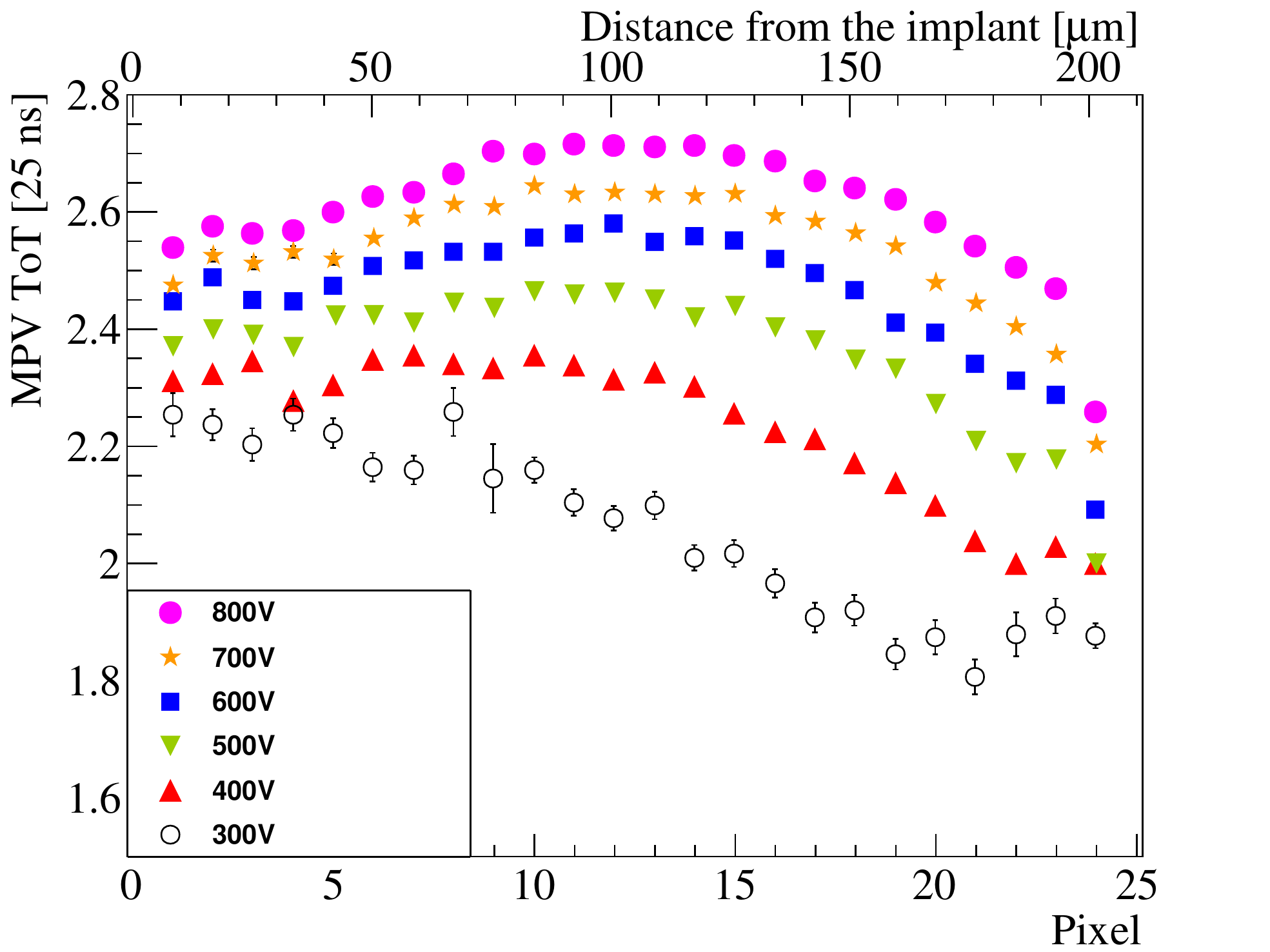}
\caption{Charge collection, expressed in units of Time over Threshold, as a function of the pixel number for a value of the cluster size equal to 24, obtained with a 
FE-I4 module, irradiated to a fluence of $2\times10^{15}$ \neqcm, employing a 200 $\mu$m thick sensor and tilted by 80$^\circ$ around its x axis with
respect to the perpendicular beam incidence. The pixel number equal to 0 corresponds to the front side and 24 to the backside of the sensor.}
\label{fig:tot_phi}
\end{figure}

\begin{figure}[hbt] 
\centering 
\includegraphics[width=\columnwidth,keepaspectratio]{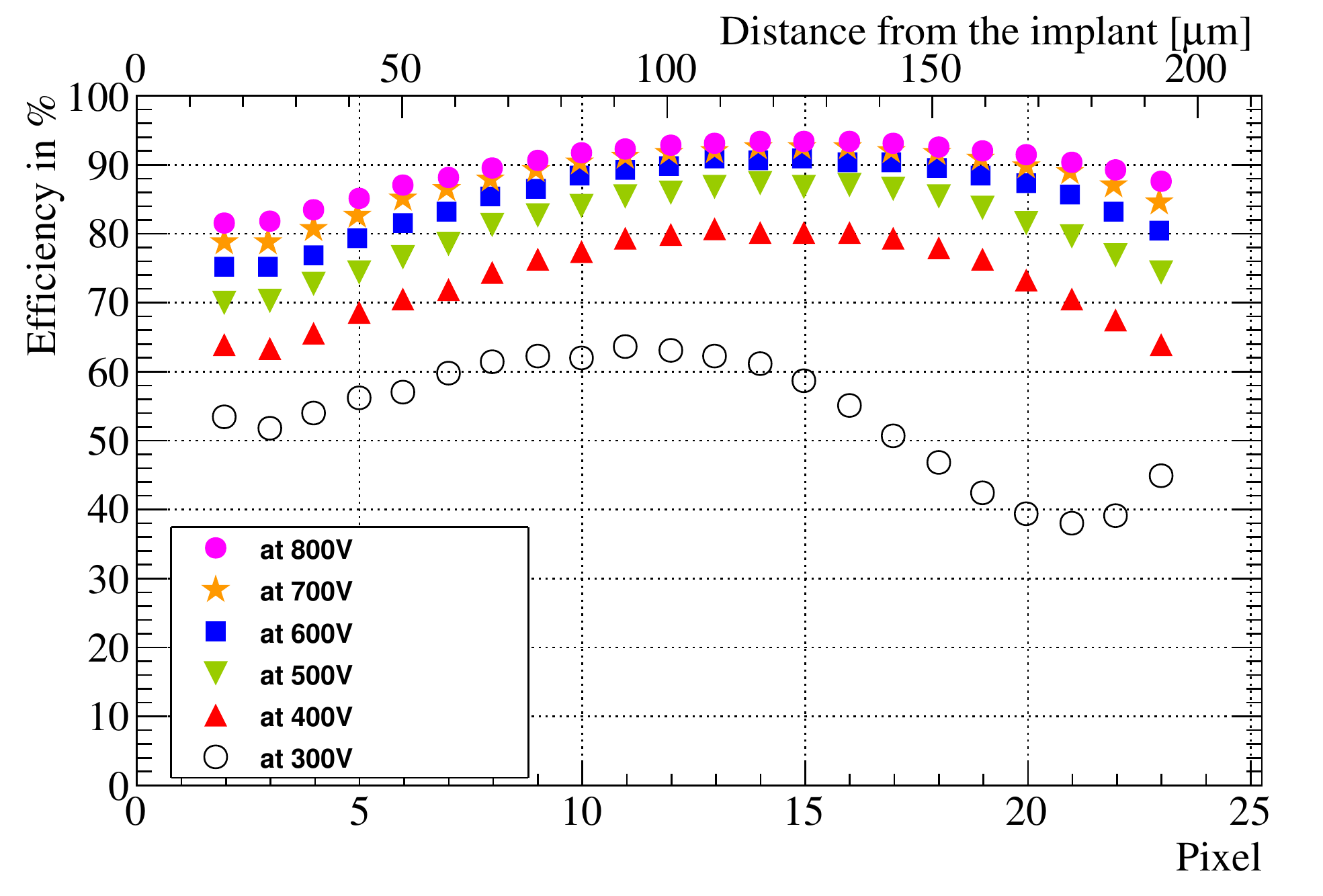}
\caption{Hit efficiency as a function of the pixel number for a value of the cluster size equal to 24, 
module employing a 200 $\mu$m thick sensor and tilted by 80$^\circ$ around its x axis with
respect to the perpendicular beam incidence. The x axis on the top indicates the corresponding value of the depth
in the silicon bulk with 0 being the frontside and 200 $\mu$m the backside.
The efficiencies of the first and the last pixels are by construction 100$\%$ since they define the cluster lenght.}
\label{fig:singlepixeleff2e15}
\end{figure}

A similar analysis was performed with a module assembled with a 200 $\mu$m thick sensor and irradiated to a fluence
of $2\times10^{15}$ \neqcm. The charge distribution in the depth of the silicon bulk is shown in Fig.\ref{fig:tot_phi},
for a bias voltage range between 300 and 800V, in the case of cluster size equal to 24 where, according to Fig.\ref{fig:clustersize} the particle crosses
all the depth of the sensor.
Pixel number 0 corresponds to the front side and pixel 24 to the backside of the sensor. It can be observed
that for lower bias voltages the collected charge decreases in the backside while at a bias voltage of 800V, resulting in a higher electric field 
throughout the bulk of the sensor, the charge collection is more uniform.
The hit efficiency for the single pixels of this irradiated module is shown in Fig.\ref{fig:singlepixeleff2e15}, also in this case only for a cluster 
size of 24. At a bias voltage of 800V, 
the range of the hit efficiency is (81.5-93.4)$\%$ at the different depths. 

\section{Conclusions}
Pixel modules assembled with thin planar sensors were investigated for the upgrade of the ATLAS pixel system at HL-LHC.
Charge collection properties and hit efficiency were analysed after irradiation for sensor thicknesses in the 
range from 100 to 200 $\mu$m.
For integrated radiation fluences of  $(4-6)\times10^{15}$ \neqcm, as expected for the second pixel layer at
HL-LHC, the best performance was observed for 100 $\mu$m thick sensors which reach the
same hit efficiency as thicker sensors already with a bias voltage of 200-300V. The highest
hit efficiency obtained for perpendicular incident tracks at these fluences is around (97-98)$\%$,
with the main inefficiency regions corresponding to the bias dot and the bias rail areas of the
pixel cell. 
New biasing structures with an external bias dot common to four pixel cells have
been investigated for thicker sensors and found to yield an higher hit efficiency, when compared to the standard design.
It is now planned to study the performance of this new layout when implemented in 100 and 150 $\mu$m thin sensors.

 Studies of cluster properties were performed for the pixel modules in the innermost layer at high
pseudorapidity for the new pixel system at HL-LHC,  where the particles traverse several pixels. In these conditions, thinner sensors
have the advantage of a lower cluster size which results in a reduced occupancy and, after irradiation, are expected
to perform better since the higher electric field counteracts trapping. 
For the ATLAS Phase II detector a smaller pitch in the z direction is foreseen, which together
with an optimal single pixel efficiency would allow to increase the precision for measuring
the entrance and the exit point of particles crossing the pixel modules at high pseudorapidity and
thus obtaining a track seed with the standalone innermost layer \cite{viel}. The performance of a 50 $\mu$m
pitch along z was therefore investigated with FE-I4 modules placed at high $\phi$ angle with respect
to the beam direction, showing a single pixel efficiency
more than 99.6$\%$ before irradiation for 100 $\mu$m thin sensors.  
A hit efficiency of (81.5-93.4$)\%$ was instead reconstructed for a module with a 200 $\mu$m thin sensor
irradiated at a fluence of $2\times10^{15}$ \neqcm. These measurements after irradiation will be also continued
with thinner sensors in a wider fluence range.

\section{Acknowledgements}
\label{sec:acknowledgment}
This work has been partially performed in the framework of the CERN RD50 Collaboration
The authors thank  V.~Cindro  for the irradiation at JSI and
A. Dierlamm for the irradiation at KIT.
Supported by the H2020 project AIDA-2020, GA no. 654168. http://aida2020.web.cern.ch/"


\bibliographystyle{model1-num-names}
\bibliography{<your-bib-database>}

\end{document}